\documentclass[aip,graphicx]{revtex4-1}
\usepackage[cp1251]{inputenc}
\usepackage[T2A]{fontenc}
\usepackage[english]{babel}
\usepackage{amssymb,latexsym,amsmath,amscd}
\usepackage{graphicx,color,framed}
\begin{document}
\selectlanguage{english}
\title{Nonlocal statistical field theory of dipolar particles forming chain-like clusters}
\author{\firstname{Yury A.} \surname{Budkov}}
\email[]{ybudkov@hse.ru}
\affiliation{G.A. Krestov Institute of Solution Chemistry of the Russian Academy of Sciences, Akademicheskaya st. 1, 153045 Ivanovo, Russia}
\affiliation{Tikhonov Moscow Institute of Electronics and Mathematics, National Research University Higher School of Economics, Tallinskaya st. 34, 123458 Moscow, Russia}
\affiliation{Landau Institute for Theoretical Physics, 142432 Chernogolovka, Russia}
\begin{abstract}
We present a nonlocal statistical field theory of a diluted solution of dipolar particles which are capable of forming chain-like clusters in accordance with the 'head-to-tail' mechanism. As in our previous study [Yu.A. Budkov 2018 J. Phys.: Condens. Matter 30 344001], we model dipolar particles as dimers comprised of oppositely charged point-like groups, separated by fluctuating distance. For the special case of the Yukawa-type distribution function of distance between the charged groups of dipolar particles we obtain an analytical expression for the electrostatic free energy of solution within the random phase approximation. We show that an increase in the association constant leads to a decrease in the absolute value of the electrostatic free energy of solution, preventing its phase separation which is in agreement with the former computer simulations and theoretical results. We obtain a non-linear integro-differential equation for the self-consistent field potential created by the fixed external charges in a solution medium, taking into account the association of dipolar particles. As a consequence of the derived self-consistent field equation, in regime of weak electrostatic interactions, we obtain an analytical expression for the electrostatic potential of the point-like test ion, surrounded by the chain-like clusters of the dipolar particles. We show that in the mean-field approximation the association does not change the bulk dielectric permittivity of the solution, but increases the solvation radius of the point-like charge, relative to the theory of non-associating dipolar particles.
\end{abstract}
\maketitle
\section{Introduction}
Up to date some milestones have been reached in the development of the theoretical models of polar fluids \cite{Teixeira2000,Tlusty2000,Onsager1936,Kirkwood1939,Hoye1976,Chandler1977,Morozov2007,Nienhuis1972,Levin1999,
Levin2001,Weiss1998,Dudowitz2004,Levy2012,Coalson1996,Abrashkin2007,
Buyukdagli2013_PRE,Buyukdagli2013_JCP,Blossey2014,Blossey2014_2,
Budkov2015,BudkovJCP2016,BudkovEA2018,McEldrew2018,Gongadze2013,Budkov2018,Martin2016,Zhuang2018}. Statistical approaches have been proposed, allowing one to calculate the static dielectric permittivity of different polar liquids \cite{Onsager1936,Kirkwood1939,Hoye1976,Chandler1977,Morozov2007,Levy2012,Zhuang2018}, describe the structural properties of polar fluids in a nano-confinement \cite{Coalson1996,Abrashkin2007,
Buyukdagli2013_PRE,Buyukdagli2013_JCP,Blossey2014,Blossey2014_2} and in the electric double layer occurring on the $"$electrolyte solution-charged electrode$"$ interfaces \cite{Budkov2015,BudkovJCP2016,BudkovEA2018,Gongadze2013,McEldrew2018}. However, currently there is a small number of theoretical approaches allowing us to calculate the free energy of polar fluid self-consistently with the dielectric permittivity \cite{Nienhuis1972,Weiss1998,Levin1999,Levin2001,Zhuang2018,Budkov2018,Martin2016}.

One of the simplest theoretical models of dipolar fluid is the model of dipolar hard spheres.
As is well known, in such a system at sufficiently low temperatures instead of a quite expectable liquid-vapour phase separation, induced by dipole-dipole interactions, it is more thermodynamically favourable to form chain-like associates \cite{Teixeira2000,Caillol1993,Weis1993,Rovigatti2011}. Up to now, the most comprehensive theoretical description of the formation of chain-like associates in the system of dipolar hard spheres has been proposed by Levin \cite{Levin1999}. The author formulated a theoretical model that allowed evaluating the free energy of dipolar hard spheres. At first, he calculated contribution of the dipole-dipole interaction to the total free energy, based on Onsager’s reaction field theory for polar fluids \cite{Onsager1936}. Based on the obtained expression for total free energy with an account for the effects of excluded volume of hard spheres and electrostatic interactions, the author obtained a "liquid-vapour" phase separation at sufficiently low temperatures. However, as it has been pointed out above, the latter contradicts the results of Monte-Carlo simulations \cite{Weis1993,Caillol1993} showing a formation of chain-like clusters, which hinder the above mentioned phase separation. Thus, Levin took into account the chain-like clusters formation, calculating the association constant in the terms of the previously proposed approach for magnetic fluids \cite{Jordan,Osipov1996}. In addition, it was supposed that electrostatic interaction contribution to the total free energy can be divided into three terms: interaction between not associated dipoles (monomers), interaction between clusters of dimension two and more, and their interaction with monomers. While the interaction of monomers was described by the above mentioned expression, the interaction of clusters with each other and with monomers was described at the Debye-Hueckel theory level. The validity of such approximation, as it seems, was justified by the physically reasonable assumption that uncompensated charges at the ends of the long enough clusters do not $"$feel$"$ the connection between them and can be considered as free ions in regular electrolyte solutions. Whereas these heuristic assumptions allowed the author to achieve the absence of phase separation at low temperatures, they cannot be justified from the first principles of statistical mechanics. Indeed, taking into account the electrostatic interactions by dividing the electrostatic free energy into three contributions that have been mentioned above appears to be a rather rough approximation. From the physical point of view, presence of chain-like associates in the system volume should mean presence of dipolar particles with different lengths. Thereby, the electrostatic interaction of clusters should be accounted for within a unified approach allowing us to describe within one formalism the thermodynamic behavior of a multi-component mixture of dipolar particles with different lengths.

Such a field-theoretic formalism has been recently proposed in our paper \cite{Budkov2018}. Instead of the conventional model of dipolar hard spheres, we have introduced a new model of dipolar particles as dimers of oppositely charged point-like groups separated by a fixed or fluctuating distance. In other words, we have attributed to each dipolar particle the arbitrary probability distribution function $g(\bold{r})$ of distance between the charged groups. Such a generalization is motivated by the fact that the recent advances in experimental studies of different organic compounds (such as proteins \cite{Canchi,Haran}, betaines \cite{Kudaibergenov,Lowe}, zwitterionic liquids \cite{Heldebrant2010}, complex colloids \cite{Wang2016}, {\sl etc}.) with charged centers, separated by a long distance from each other (about $ 5-20~nm$) require formulating analytical and numerical approaches taking into account the molecules internal electric structure \cite{Shen2016,van Blaaderen,Dussi2013}. Within the previously proposed nonlocal field-theoretical model, we have derived a non-linear integro-differential equation with respect to the mean-field electrostatic potential, generalizing the Poisson-Boltzmann-Langevin equation for the point-like dipoles obtained in papers \cite{Abrashkin2007,Coalson1996} and generalized for the case of a polarizable solvent in \cite{Budkov2015}. We have applied the obtained equation in its linearized form to derive expressions for the mean-field electrostatic potential of the point-like test ion and its solvation free energy in a salt-free solution, as well as in a solution in the presence of salt ions. Moreover, we have obtained a general expression for the bulk electrostatic free energy of the ion-dipole mixture within the Random phase approximation (RPA).

In the current study we would like to present a generalization of our previous theory \cite{Budkov2018} for the case of formation of chain-like clusters from dipolar particles, according to the $"$head-to-tail$"$ mechanism \cite{Levin1999,Levin2001,Dudowitz2004}. More specifically, the following two questions will be addressed in this paper:

\begin{itemize}
\item{How does the formation of the chain-like clusters influence on the thermodynamic stability of the solutions of dipolar particles?}

\item{How will the electrostatic potential of point-like test ion, surrounded by the chain-like clusters, built from the dipolar particles differ from the potential of point-like test ion immersed to solution of non-associating dipolar particles?}
\end{itemize}
The rest of the article is organized as follows. In Sec. II we present the general theory of diluted solution of dipolar particles that are capable of forming chain-like clusters in the presence of external charges with a fixed density in the system volume. Sec. III shows how the formulated theory can be applied to description of a bulk solution in the absence of external charges in the system volume. Sec. IV presents a calculation of the electrostatic potential of a point-like test ion in the environment of the chain-like associates. Sec. V contains conclusions and a description of some further perspectives.

\section{General theory}
Let us consider a diluted solution of dipolar particles capable of forming chain-like clusters, according to the $"$head-to-tail$"$ mechanism. Let us assume that the chain-like clusters are in associative equilibrium at the temperature $T$ and are confined in volume $V$. For simplicity, we suppose that association of two oppositely charged groups with charges $\pm e$ ($e$ is the elementary charge) of different dipolar particles leads to the formation of an electrically neutral 'linker' not participating in the electrostatic interactions anymore. As in our previous paper \cite{Budkov2018}, we attribute with each dipolar particle a probability distribution function $g(\bold{r})$ of the distance between its charged centers. We also assume that the dipolar particles are dissolved in some low-molecular solvent, which we model as a continuous dielectric medium with the dielectric permittivity $\varepsilon$. Thereby, within our model, the solution of dipolar particles is modelled as a set of chain-like clusters of different dimensions $n$ with oppositely charged end groups, composed of $n$ dipolar particles immersed in a continuous dielectric background. Moreover, we attribute to each appearing bond the association constant $K$. In general case, the association constant can be written as $K=v\exp\left[-\Delta f/k_{B}T\right]$, where $v$ is the so-called configurational volume \cite{van Roij1996}, whereas $\Delta f$ is the free energy of association \cite{Dudowitz2004,Dudowitz2000_2,Dudowitz2000_1}, $k_B$ is the Boltzmann constant. Physically, the configuration volume determines the effective volume, in which the bonded oppositely charged groups are localized. The free energy of association $\Delta f=\Delta \epsilon-T\Delta s$ depends on the chemical specificity of dipolar particles and is generally determined by the energy $\Delta \epsilon$ of electrostatic interactions (that is usually the main contribution), specific interactions, such as hydrogen bonding, pi-pi stacking, {\sl etc}, and configuration entropy $\Delta s$, related to the internal degrees of freedom of the associated groups. In the general case, both variables $v$ and $\Delta f$ must be the functions of temperature \cite{van Roij1996,Dudowitz2004}. However, in the present study to investigate the basic effects, related to the association of dipolar particles, we will not specify the contributions to association constant $K$ and its temperature dependence, but consider it as a phenomenological parameter.

We consider only the case of a quite diluted solution of the dipolar particles, neglecting all intermolecular interactions except the electrostatic ones. In other words, due to the fact that in this research we would like to study only the effects of chain-like clusters formation and electrostatic interactions, we will neglect, for simplicity, the dispersion and excluded volume interactions contribution to the total free energy. The simplest way to account for these universal intermolecular interactions in the thermodynamic behavior description of the solution, is to add additional terms to the total free energy, according to the Van der Waals \cite{van Roij1996}, Carnahan-Starling \cite{Levin1999,Levin2001}, or Flory-Huggins \cite{Dudowitz2004} equations of state. As it has been already pointed out in the Introduction, in this study we focus on the effects of formation of the chain-like clusters, so that we do not take into account the ring and branched structures which can also take place in the dipolar fluids \cite{Tlusty2000,Dussi2013}. In order to account for the branched and ring structures, one can use the sophisticated approach, developed in works \cite{Erukhimovich1995,Erukhimovich1999,Erukhimovich2002}.

Therefore, the total Helmholtz free energy of a diluted solution of dipolar particles can be written as
\begin{equation}
\label{Ftot}
F(\{\rho_{n}\})=F_{id}(\{\rho_{n}\})+F_{el}(\{\rho_{n}\}),
\end{equation}
where
\begin{equation}
\label{Fid1}
F_{id}(\{\rho_{n}\})=Vk_{B}T\sum\limits_{n=1}^{\infty}\rho_{n}\left(\ln{\rho_{n}}-1\right)-k_{B}T\ln K^{\mathcal{M}(\{\rho_{n}\})}
\end{equation}
is the ideal free energy of the chain-like clusters \cite{van Roij1996}, where the total number of bonds can be calculated as
\begin{equation}
\label{bonds_numb}
\mathcal{M}(\{\rho_{n}\})=V\sum\limits_{n=1}^{\infty}\rho_{n}(n-1),
\end{equation}
where $\rho_{n}$ is the number density of clusters of dimension $n$.  We would like to point out that the total number density $\rho$ of the dipolar particles can be calculated from the mass conservation law
\begin{equation}
\label{tot_density}
\rho = \sum\limits_{n=1}^{\infty} n\rho_{n}.
\end{equation}

The contribution of electrostatic interactions can be calculated in the standard way:
\begin{equation}
\label{elect}
F_{el}=-k_{B}T\ln{Q},
\end{equation}
where
\begin{equation}
Q=\int d\Gamma \exp\left[-\beta H\right]
\end{equation}
is the configuration integral with the Hamiltonian of electrostatic interactions
\begin{equation}
H=\frac{1}{2}\int d\bold{r} \int d\bold{r}^{\prime}\hat{\rho}(\bold{r})G_{0}(\bold{r}-\bold{r}^{\prime})\hat{\rho}(\bold{r}^{\prime})=\frac{1}{2}(\hat{\rho}G_0 \hat{\rho}),
\end{equation}
where $G_{0}(\bold{r}-\bold{r}^{\prime})=1/(4\pi\varepsilon\varepsilon_0 |\bold{r}-\bold{r}^{\prime}|)$ is the Green function of the Poisson equation; $\beta=1/{k_{B}T}$ is the inverse thermal energy with the Boltzmann constant $k_B$ and the temperature $T$; the local charge density of the system can be written as
\begin{equation}
\hat{\rho}(\bold{r})=e\sum\limits_{n=1}^{\infty}\sum\limits_{j=1}^{N_{n}}\left(\delta(\bold{r}-\bold{r}^{(+)}_{n,j})-\delta(\bold{r}-\bold{r}^{(-)}_{n,j})\right)
+\rho_{ext}(\bold{r}),
\end{equation}
where $\bold{r}^{(\pm)}_{n,j}$ are the coordinates of the charged end groups of the chain-like clusters, $e$ is the elementary charge, $N_n$ is the number of clusters of dimension $n$; $\rho_{ext}(\bold{r})$ is the density of external charges. The integration measure over the coordinates of the chain-like clusters can be written as
\begin{equation}
\label{int_meas}
\int d\Gamma(\cdot)=\int..\int\prod\limits_{n=1}^{\infty} d\Gamma_{n}(\cdot),
\end{equation}
where
\begin{equation}
\label{int_meas}
\int d\Gamma_n(\cdot)=\int\frac{d\bold{r}d\bold{r}^{\prime}}{V}g_{n}(\bold{r}-\bold{r}^{\prime})(\cdot),
\end{equation}
and
\begin{equation}
g_{n}(\bold{r}-\bold{r}^{\prime})=\int d\bold{r}_{1}\int d\bold{r}_{2}..\int d\bold{r}_{n-1}g(\bold{r}-\bold{r}_{1})g(\bold{r}_1-\bold{r}_{2})..
g(\bold{r}_{n-1}-\bold{r}^{\prime})
\end{equation}
is the probability distribution function of distance between the charged ends of $n$-mer.

Further, using the standard Hubbard-Stratonovich transformation
\begin{equation}
\exp\left[-\frac{\beta}{2}(\hat\rho G_0\hat\rho)\right]=\int\frac{\mathcal{D}\varphi}{C}\exp\left[-\frac{\beta}{2}(\varphi G_0^{-1}\varphi)+i\beta(\hat\rho\varphi)\right],
\end{equation}
we arrive at the following functional representation of the configuration integral
\begin{equation}
Q=\int\frac{\mathcal{D}\varphi}{C}\exp\left[-\frac{\beta}{2}(\varphi G_0^{-1}\varphi)+i\beta(\rho_{ext}\varphi)\right]\prod\limits_{n=1}^{\infty}Q_n^{N_n}[\varphi]
\end{equation}
with the one-cluster partition functions
\begin{equation}
Q_n[\varphi]=\int\frac{d\bold r^{(+)}d\bold r^{(-)}}{V}g_{n}(\bold r^{(+)}-\bold r^{(-)})\exp(i\beta e(\varphi(\bold r^{(+)})-\varphi(\bold r^{(-)})))
\end{equation}
and the following short-hand notations
$$(\varphi G_0^{-1}\varphi)=\int d\bold r\int d\bold r'\varphi(\bold r)G_0^{-1}(\bold r,\bold r')\varphi(\bold r'),~~(f\varphi)=\int d\bold r f(\bold r)\varphi(\bold r)$$
and
$$C=\int \mathcal{D}\varphi\exp\left[-\frac{\beta}{2}(\varphi G_0^{-1}\varphi)\right]$$
is the normalization constant of the Gaussian measure. The reciprocal Green function $G_0^{-1}$ is determined by the following integral relation
\begin{equation}
\int d \bold{r}'' G_0(\bold r-\bold r'')G_0^{-1}(\bold r'',\bold r')=\delta(\bold r-\bold r').
\end{equation}

In the thermodynamic limit,
$V\to \infty,~N_{n}\to\infty, N_{n}/V\to \rho_{n}$,
we obtain \cite{Efimov1996,Budkov2018}
\begin{equation}
\nonumber
Q_n^{N_n}[\varphi]=\left[1+\frac{1}{V}\int d\bold r^{(+)}\int d\bold r^{(-)}g_n(\bold r^{(+)}-\bold r^{(-)})(\exp\left(i\beta e(\varphi(\bold r^{(+)})-\varphi(\bold  r^{(-)}))\right)-1)\right]^{N_n}\simeq
\end{equation}
\begin{equation}
\exp\left[\rho_n\int d\bold r^{(+)}\int d\bold r^{(-)}g_n(\bold r^{(+)}-\bold r^{(-)})(\exp\left(i\beta e(\varphi(\bold r^{(+)})-\varphi(\bold  r^{(-)}))\right)-1)\right].
\end{equation}
Thereby, we arrive at the following functional representation of the configuration integral
\begin{equation}
\label{func_int}
Q=\int\frac{\mathcal{D}\varphi}{C}\exp\left[-S[\varphi]\right],
\end{equation}
where the following functional
\begin{equation}
S[\varphi]=\frac{\beta}{2}(\varphi G_0^{-1}\varphi)-i\beta(\rho_{ext}\varphi)-\int d\bold r\int d\bold r'\mathcal{C}(\bold r-\bold r')\left(\exp\left[i\beta e(\varphi(\bold r)-\varphi(\bold r'))\right]-1\right)
\end{equation}
with the following auxiliary kernel
\begin{equation}
\mathcal{C}(\bold{r}-\bold{r}^{\prime})=\sum\limits_{n=1}^{\infty}\rho_ng_{n}(\bold{r}-\bold{r}^{\prime}).
\end{equation}

In order to obtain the electrostatic free energy of the dipolar fluid within the Random phase approximation (RPA), we expand the functional $S[\varphi]$ in (\ref{func_int}) into a power series near the mean-field $\varphi(\bold{r})=i\psi(\bold{r})$ and truncate it by the quadratic term, i.e.
\begin{equation}
S[\varphi]\approx S[i\psi]+\frac{\beta}{2}\left(\varphi\mathcal{G}^{-1}\varphi\right),
\end{equation}
where
\begin{equation}
\nonumber
\mathcal{G}^{-1}(\bold r,\bold r'|\psi)=k_{B}T\frac{\delta^2 S[i\psi]}{\delta \varphi(\bold{r})\delta \varphi(\bold{r}')}
\end{equation}
\begin{equation}
\nonumber
=G_0^{-1}(\bold r,\bold r')+\frac{2 e^2}{k_BT}\int d\bold{r}''\mathcal{C}(\bold{r}-\bold{r}'')\cosh\left(\frac{e\left(\psi(\bold{r})-\psi(\bold{r}'')\right)}{k_{B}T}\right)(\delta(\bold r-\bold r')-\delta(\bold r''-\bold r'))
\end{equation}
is the renormalized reciprocal Green function with the mean-field electrostatic potential $\psi(\bold{r})$ satisfying the mean-field equation $\delta S/\delta\varphi(\bold{r})=0$, which can be written as follows
\begin{equation}
\label{mean-field_eq}
-\varepsilon\varepsilon_0 \Delta \psi(\bold{r})=2e\int d\bold{r}^{\prime}\mathcal{C}(\bold{r}-\bold{r}^{\prime})\sinh\frac{e(\psi(\bold{r})-\psi(\bold{r}^{\prime}))}{k_{B}T}+\rho_{ext}(\bold{r}).
\end{equation}
Note that equation (\ref{mean-field_eq}) is a generalization of the early obtained self-consistent field equation \cite{Budkov2018} for the case of associating dipolar particles.

Therefore, taking the Gaussian functional integral \cite{Podgornik1989}, we obtain the following general relation for the configuration integral in the RPA
\begin{equation}
\label{RPA_gen}
Q\approx \exp\left\{-S[i\psi]+\frac{1}{2}tr\left(\ln \mathcal{G} -\ln G_{0}\right)\right\},
\end{equation}
where the symbol $tr(..)$ means the trace of the integral operator \cite{Netz2001,Podgornik1989,Lue2006}.

Thus, electrostatic free energy take the form:
\begin{equation}
F_{el}=k_{B}T\left(S[i\psi]-\frac{1}{2}tr\left(\ln \mathcal{G} -\ln G_{0}\right)\right).
\end{equation}

\section{Bulk solution theory}
Now we will consider the case when there are no external charges in the system volume, i.e. $\rho_{ext}(\bold{r})=0$. Thus, in that case the mean-field electrostatic potential $\psi(\bold{r})=0$, so that $S[0]=0$, and the electrostatic free energy is determined only by the thermal fluctuations of the electrostatic potential near its zero value \cite{Budkov2018,Podgornik1989} and can be calculated by
\begin{equation}
\label{Fcor_ass}
F_{el}(\{\rho_{n}\})=\frac{Vk_{B}T}{2}\int\frac{d\bold{k}}{(2\pi)^3}
\left(\ln\left(1+\frac{\varkappa^2(\bold{k};\{\rho_{n}\})}{k^2}\right)-\frac{\varkappa^2(\bold{k};\{\rho_{n}\})}{k^2}\right),
\end{equation}
where the screening function is
\begin{equation}
\label{scr_func}
\varkappa^2(\bold{k};\{\rho_{n}\})=\frac{2e^2}{\varepsilon\varepsilon_0 k_{B}T}\sum\limits_{n=1}^{\infty}\rho_{n}\left(1-g^{n}(\bold{k})\right).
\end{equation}
We would also like to stress that we have subtracted the electrostatic self-energy of the chain-like clusters from the final expression of the electrostatic free energy \cite{Borue1988}.

At the thermodynamic equilibrium the following relation has to be satisfied
\begin{equation}
\label{equil_cond1}
\mu_{n}=\frac{1}{V}\frac{\partial{F}}{\partial{\rho_{n}}}=n\mu_{1}
\end{equation}
which yields
\begin{equation}
\label{equil_cond2}
\rho_{n}=K^{n-1}\rho_{1}^{n}\exp\left[\frac{n\mu_{1}^{(el)}-\mu_{n}^{(el)}}{k_{B}T}\right],
\end{equation}
where the electrostatic chemical potential of monomers
\begin{equation}
\label{mon_chem_pot}
\frac{\mu_{1}^{(el)}}{k_{B}T}=\frac{1}{Vk_{B}T}\frac{\partial F_{el}}{\partial \rho_{1}}=-\frac{e^2}{\varepsilon\varepsilon_0 k_{B}T}\int\frac{d\bold{k}}{(2\pi)^3}\frac{1-g(\bold{k})}{k^2}\frac{\varkappa^2(\bold{k};\{\rho_{n}\})}{k^2+\varkappa^2(\bold{k};\{\rho_{n}\})}
\end{equation}
is introduced. The electrostatic chemical potentials of $n$-mers ($n\geq 2$) can be estimated by
\begin{equation}
\nonumber
\frac{\mu_{n}^{(el)}}{k_{B}T}=\frac{1}{Vk_{B}T}\frac{\partial F_{el}}{\partial \rho_{n}}=-\frac{e^2}{\varepsilon\varepsilon_0 k_{B}T}\int\frac{d\bold{k}}{(2\pi)^3}\frac{1-g^{n}(\bold{k})}{k^2}\frac{\varkappa^2(\bold{k};\{\rho_{n}\})}{k^2+\varkappa^2(\bold{k};\{\rho_{n}\})}=
\end{equation}
\begin{equation}
\nonumber
-\frac{e^2}{\varepsilon\varepsilon_0 k_{B}T}\int\frac{d\bold{k}}{(2\pi)^3}\frac{(1-{g}(\bold{k}))\left(1+\sum\limits_{k=1}^{n-1}g^{k}(\bold{k})\right)}{k^2}
\frac{\varkappa^2(\bold{k};\{\rho_{n}\})}{k^2+\varkappa^2(\bold{k};\{\rho_{n}\})}\approx
\end{equation}
\begin{equation}
\label{nmer_chem_pot}
-\frac{e^2n}{\varepsilon\varepsilon_0 k_{B}T}\int\frac{d\bold{k}}{(2\pi)^3}\frac{1-{g}(\bold{k})}{k^2}
\frac{\varkappa^2(\bold{k};\{\rho_{n}\})}{k^2+\varkappa^2(\bold{k};\{\rho_{n}\})}=n\mu_{1}^{(el)},
\end{equation}
so that $n\mu_{1}^{(el)}-\mu_{n}^{(el)}\approx 0$.
Therefore, in this approximation we neglect the influence of electrostatic interactions of the charged end groups of the chain-like clusters on the associative equilibrium, assuming that
\begin{equation}
\label{equil_cond3}
\rho_{n}\approx K^{n-1}\rho_{1}^{n}.
\end{equation}

Substituting (\ref{equil_cond3}) with the relation (\ref{tot_density}), we arrive at the equation
\begin{equation}
\rho = \frac{\rho_{1}}{\left(1-K\rho_{1}\right)^2}
\end{equation}
that yields
\begin{equation}
\label{rho_mon}
\rho_{1}=\frac{1}{2K^2\rho}\left(1+2K\rho-\sqrt{1+4K\rho}\right).
\end{equation}
The screening function can also be easily calculated, giving
\begin{equation}
\label{scren_funct}
\varkappa^2(\bold{k};\{\rho_{n}\})=\kappa_{D}^2(1-K\rho_{1})\frac{1-g(\bold{k})}{1-K\rho_{1}g(\bold{k})},
\end{equation}
where $\kappa_{D}=\left(2e^2\rho/\varepsilon\varepsilon_0 k_{B}T\right)^{1/2}$ is the inverse Debye radius, associated with the charged groups of the dipolar particles in the absence of dipole association \cite{Budkov2018}.

Further, using relations (\ref{equil_cond3}) and (\ref{rho_mon}) and the relation for the characteristic function
\begin{equation}
\label{charact_func}
g(\bold{k})=\frac{1}{1+\frac{k^2l^2}{6}},
\end{equation}
determining the Yukawa-type probability distribution function \cite{Budkov2018,Gordievskaya2018}, after simple calculations, we arrive at
\begin{equation}
F=F_{id}+F_{el},
\end{equation}
where
\begin{equation}
\label{Fid2}
F_{id}=V\rho k_{B}T\left(\ln\left(\frac{\sqrt{1+4K\rho}-1}{\sqrt{1+4K\rho}+1}\right)-\frac{\sqrt{1+4K\rho}-1}{2K\rho}\right)
\end{equation}
is the free energy of the ideal solution of chain-like clusters \cite{van Roij1996} and
\begin{equation}
\label{Fcor_ass2}
F_{el}=-\frac{Vk_{B}T}{l^3}\frac{\sigma\left(y\right)}{\bar{n}^{3/2}(K\rho)},
\end{equation}
is the electrostatic free energy with the auxiliary function
\begin{equation}
\sigma(y)=\frac{\sqrt{6}}{4\pi}(2(1+y)^{3/2}-2-3y)
\end{equation}
and $y=\kappa_{D}^2l^2/6$ is the strength of the dipole-dipole interactions \cite{Budkov2018}; $l^2/6$ is the mean-square distance between charged centers of dipolar particles;
\begin{equation}
\label{average}
\bar{n}(K\rho)=\frac{\sum\limits_{n=1}^{\infty}n\rho_n}{\sum\limits_{n=1}^{\infty}\rho_n}=\frac{1+\sqrt{1+4K\rho}}{2}
\end{equation}
is the average number of dipolar particles in the chain-like clusters. As one can see, an increase in the association constant results in an expectable decrease in the absolute value of electrostatic free energy of solution. Fig. 1 demonstrates the dependences of osmotic pressure $\Pi =-\left(\partial{F}/\partial{V}\right)_{T}$ of a solution as a function of the concentration of dipolar particles, presented for different values of the association constant. At the zero association constant and at sufficiently big concentrations (at all temperatures without accounting for the excluded volume interactions), electrostatic interactions of dipolar particles lead to the homogeneous phase ceasing to be stable. In order to obtain the van der Waals loop on the osmotic pressure curve, from which one can calculate concentrations in the coexisting phases, it is necessary to take into account the excluded volume interactions \cite{van Roij1996}. As is seen, an increase in the association constant stabilizes the solution and at its rather big value the region of unstable states (where $\partial{\Pi}/\partial{\rho}< 0$) disappears. Note that increase in the association constant leads to an expectable decrease in the osmotic pressure.

\begin{figure}[h!]
\center{\includegraphics[width=0.8\linewidth]{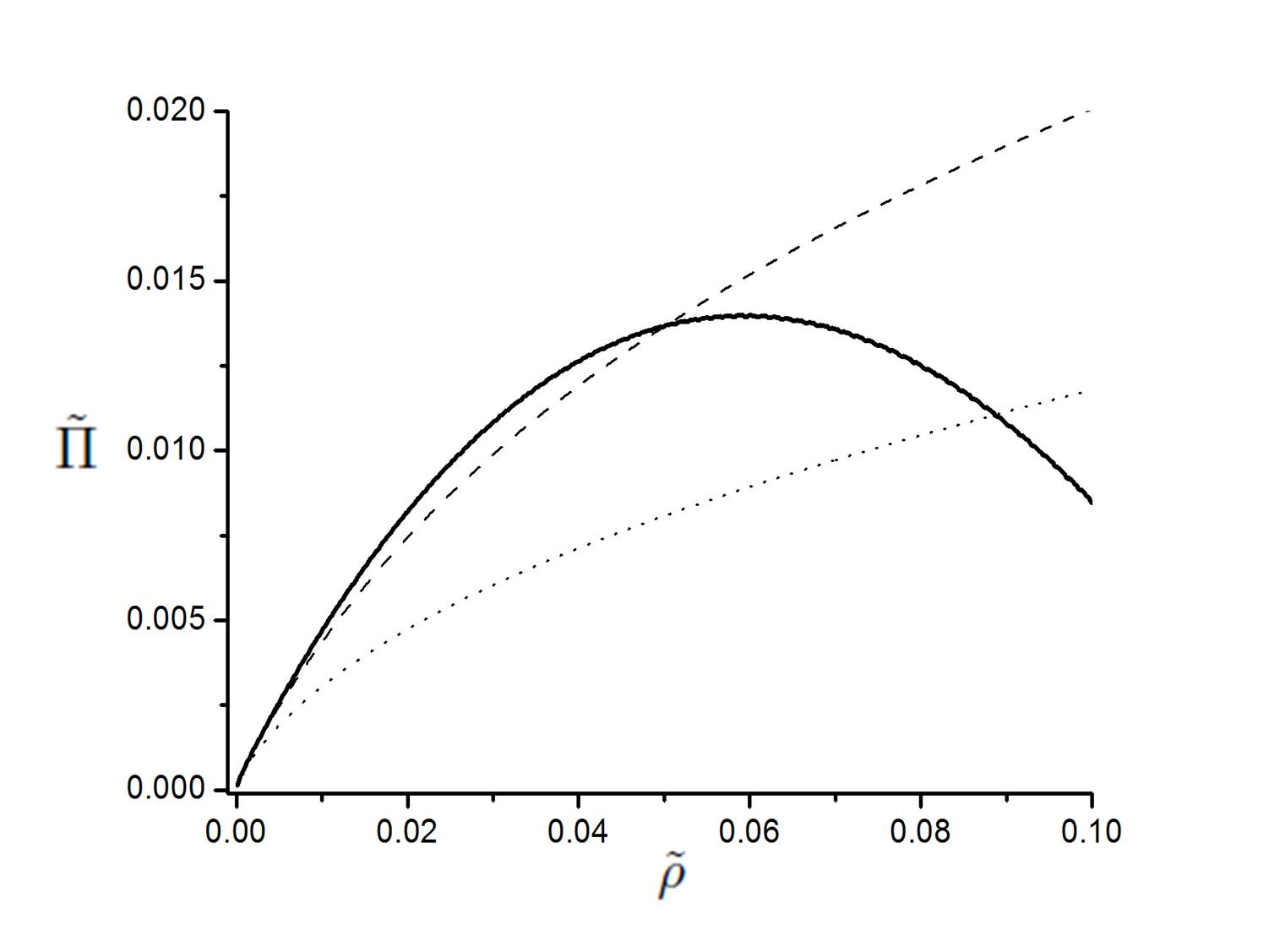}}
\caption{Reduced osmotic pressure $\tilde{\Pi}=(4\pi \varepsilon\varepsilon_{0}\Pi l^4)/e^2$ as a function of reduced concentration of dipolar particles $\tilde{\rho}=\rho l^3$, plotted for different reduced association constants $\tilde{K}=Kl^{-3}=0$ (solid line), $\tilde{K}=10$ (dashed line) and $\tilde{K}=100$ (dotted line) and fixed reduced temperature $\tilde{T}=(4\pi \varepsilon\varepsilon_{0}k_{B}T l)/e^2$=0.5.}
\label{fig1}
\end{figure}

Expression (\ref{Fcor_ass2}) for the electrostatic free energy can be obtained from the following scaling arguments. From the scaling point of view, at sufficiently small concentrations, for which contribution of excluded volume interactions is small, the electrostatic free energy density of the non-associating dipolar particles can be written as follows
\begin{equation}
\frac{f_{el}}{k_B T}=l^{-3}s\left(l/r_{D}\right),
\end{equation}
where $s(x)$ is some dimensionless universal function. For the case of associating dipolar particles, taking into account that dipolar clusters considered to be long dipolar particles, we can write
\begin{equation}
\frac{\tilde{f}_{el}}{k_B T}=\tilde{l}^{-3}s\left(\tilde{l}/\tilde{r}_{D}\right),
\end{equation}
where $\tilde{f}_{el}$ is the electrostatic free energy density of the solution with association, $\tilde{l}$ is the average size of the chain-like cluster and $\tilde{r}_{D}=(2e^2\rho_c/\varepsilon\varepsilon_0 k_{B}T)^{-1/2}$ is the Debye radius, attributed to the charged groups of the chain-like clusters with their total concentration $\rho_c=\sum_{n=1}^{\infty}\rho_n$. Further, taking into account that for rather long chain-like clusters ($\bar{n}\gg 1$), in accordance with the central limit theorem \cite{Gnedenko}, $\tilde{l}\simeq l\sqrt{\bar{n}}$ and that $\rho_{c}\approx \rho/\bar{n}$, we arrive at the relation $\tilde{l}/\tilde{r}_D\approx l/r_{D}$. Thereby, we obtain
\begin{equation}
\tilde{f}_{el}=\frac{f_{el}}{\bar{n}^{3/2}}.
\end{equation}
Therefore, using the expression for the electrostatic free energy of fluid of non-associating dipolar particles (see, eq. (61) in \cite{Budkov2018}), we arrive at (\ref{Fcor_ass2}).

We would like to note that the presented theoretical results must be valid at sufficiently large association constants, for which even at sufficiently small dipolar particles concentrations the average dimension of associates $\bar{n}\gg 1$. Indeed, only in that case we can safely neglect the influence of electrostatic interactions between clusters on the associative equilibrium. As it follows from the relation (\ref{average}), the latter condition can be satisfied at $K\gg 1/\rho$. We would like to note that if the latter condition satisfied, then one can consider the dimension of clusters $n$ as continual variable distributed in accordance with the exponential probability distribution function $p(n)\simeq (K\rho)^{-1/2}\exp\left[-n/(K\rho)^{1/2}\right]$ with the average $\bar{n}\simeq (K\rho)^{1/2}$.

\section{Point-like charge in dipole environment: effect of dipole association}
Now we would like to understand how chain-like clusters formation of dipolar particles affects the solvation quantities of the point-like test ion carrying charge $q$, immersed to solution. Especially, it is interesting to calculate the electrostatic potential profile of test ion in dipolar particles solution medium. In order to make this, we must solve self-consistent field equation (\ref{mean-field_eq}), taking into account that $\rho_{ext}(\bold{r})=q\delta(\bold{r})$. In order to obtain the analytical expression for the electrostatic potential, we consider the case of weak electrostatics, assuming that $e\psi/k_{B}T\ll 1$. In that case we can linearize equation (\ref{mean-field_eq}), that yields
\begin{equation}
\label{mean-field_eq_lin}
-\varepsilon\varepsilon_0 \Delta \psi(\bold{r})=\frac{2e^2}{k_{B}T}\sum\limits_{n=1}^{\infty}\rho_{n}\int d\bold{r}^{\prime}\left(\delta(\bold{r}-\bold{r}^{\prime})-g_{n}(\bold{r}-\bold{r}^{\prime})\right)\psi(\bold{r}^{\prime})+q\delta(\bold{r}).
\end{equation}

Using the Fourier-transformation \cite{Hormander}, we obtain the solution of equation (\ref{mean-field_eq_lin}) in the standard linear-response theory form \cite{Lue2006,Victorov2016}
\begin{equation}
\label{solution}
\psi(\bold{r})=\frac{q}{\varepsilon\varepsilon_{0}}\int\frac{d\bold{k}}{(2\pi)^3}\frac{ \exp\left(i\bold{k}\bold{r}\right)}{k^2+\varkappa^2(\bold{k};\{\rho_{n}\})},
\end{equation}
where the screening function $\varkappa^2(\bold{k};\{\rho_{n}\})$ is determined by expression (\ref{scr_func}). Further, assuming that in the limit of weak electrostatic interactions the presence of the test charge does not change the associative equilibrium, relative to the one realized in the bulk solution and, consequently, using expression (\ref{scren_funct}) for the screening function with characteristic function (\ref{charact_func}), we obtain the following analytical expression for the electrostatic potential
\begin{equation}
\label{potential_fin}
\psi(\bold{r})=\frac{q}{4\pi\varepsilon\varepsilon_0r}\frac{1+y\exp\left(-\frac{r}{l_s}\right)}{1+y},
\end{equation}
where $l_s$ is the effective solvation radius of the point-like test ion, determined by the relation
\begin{equation}
\label{solv_rad}
l_{s}=\frac{l}{\sqrt{6\varepsilon(1+y)}}\bar{n}^{1/2}(K\rho).
\end{equation}
As one can see from equation (\ref{potential_fin}), in the mean-field approximation, accounting for the chain-like clusters formation of the dipolar particles does not change the bulk dielectric permittivity being \cite{Budkov2018}
\begin{equation}
\label{diel_perm}
\varepsilon_b = \varepsilon(1+y),
\end{equation}
but leads to the renormalization of the solvation radius $l_{s}$. Indeed, as it can be seen from (\ref{solv_rad}), an increase in association constant results in the wider region, where the local dielectric permittivity
\begin{equation}
\varepsilon(r)=\frac{\varepsilon_b}{1+y\exp\left(-\frac{r}{l_s}\right)}
\end{equation}
is smaller than it is in the bulk solution.

We can also calculate as follows the effective solvation free energy of the point-like test ion \cite{Budkov2018}  in the linear approximation
\begin{equation}
\label{solv}
\Delta F_{solv}=\frac{1}{2}\int d\bold{r}\rho_{ext}(\bold{r})\left(\psi(\bold{r})-\psi_{ext}(\bold{r})\right)=-\frac{q^2\sqrt{6}}{8\pi\varepsilon\varepsilon_0 l}\frac{y}{\sqrt{1+y}}\bar{n}^{-1/2}(K\rho),
\end{equation}
where $\psi_{ext}(\bold{r})=q/\left(4\pi\varepsilon\varepsilon_{0}r\right)$ is the potential of the point-like ion in the pure solvent.

We would like to note that like expression (\ref{Fcor_ass2}) for electrostatic free energy, expressions (\ref{solv_rad}), (\ref{diel_perm}) and (\ref{solv}) can be obtained from the scaling arguments given in section III.

\section{Concluding remarks and perspectives}
In conclusion, we have formulated a nonlocal statistical field theory of the diluted solution of chain-like clusters, formed from dipolar particles according to the head-to-tail mechanism. Using the field-theoretic formalism developed in our previous paper \cite{Budkov2018}, we have calculated the bulk electrostatic free energy of the chain-like cluster. We have shown that an increase in the association constant leads to a decrease in the absolute value of the electrostatic free energy, preventing a liquid-vapour phase separation, induced by electrostatic correlation attraction. We have derived the non-linear integro-differential equation for the electrostatic potential of self-consistent field, generated by external electric charges in the solution medium and charged end groups of chain-like clusters, taking into account the association of dipolar particles. As an application of the obtained self-consistent field equation, in the regime of weak electrostatic interactions, we have derived an analytical expression for the electrostatic potential of a point-like test ion surrounded by chain-like clusters of different lengths. We have established that in the mean-field approximation the association does not change the dielectric permittivity of the bulk solution, but affects significantly the solvation radius of the test point-like ion. In other words, we have found that an increase in the association constant makes the region around the test point-like ion wider, with the local dielectric permittivity being smaller than that in the bulk phase.

As it was pointed out in the main text, in the present study we have neglected the contributions of excluded volume effect of dipolar particles. This assumption is motivated by the fact that we considered only a rather diluted solution of dipolar particles. In order to account for the excluded volume effect, we can describe the dipolar particles as hard dumbbells. In that case, we can use the virial equation of state for hard dumbbells, as it was made in a recent paper \cite{Gordievskaya2018}. On the other hand, in a system of charged dumbbells at a rather low temperature apart from the chain-like clusters ring and branched associates can be also formed \cite{Dussi2013}. Thus, in order to describe the phase behavior of the associated charged dumbbells fluid, apart from the excluded volume and electrostatic interactions, it is necessary to take into account the associative equilibrium of chain-like, ring, and branched clusters. For this purpose, one can use the formalism, proposed in papers \cite{Erukhimovich1995,Erukhimovich1999,Erukhimovich2002}. We would like to note that the formulated theory could be relevant for the solutions of complex dipolar patchy colloid particles  consisting of nanoparticles with adhesive functional molecular groups, such as proteins \cite{Shen2016,van Blaaderen}.

\begin{acknowledgements}
Development of the model of dipolar particles association was supported by the Russian Science Foundation project No. 18-71-10061. The results, presented in sec. IV were funded by RFBR according to the research project No 18-31-20015. The author thanks I.Ya. Erukhimovich for the fruitful and motivating discussions.
\end{acknowledgements}

\end{document}